# *Artificial Intelligence for Inclusive Engineering Education: Advancing Equality, Diversity, and Ethical Leadership*

Mona G. Ibrahim
*School of Energy and Environmental Engineering*
*Egypt-Japan University of Science and Technology(E-JUST)*
Alexandria , Egypt
mona.gamal@ejust.edu.eg

Riham Hilal
*Faculty of Art & Design*
*Egypt-Japan University of Science and Technology(E-JUST)*
Alexandria , Egypt
riham.hilal@ejust.edu.eg

*Abstract*— AI technology development has transformed the field of engineering education with its adaptivity-driven, data-based, and ethical-led learning platforms that promote equity, diversity, and inclusivity. But with so much progress being made in so many areas, there are unfortunately gaps in gender equity, representation in cultures around the world, and access to education and jobs in stem education. The paper describes an ethical approach to using AI technology that supports the United Nations 2030 agenda for sustainability. In particular, this includes both Goal 5--Gender Equity--and Goal 10--Reducing Inequalities. Based on a synthesis strategy using both critical thinking strategies related to case studies around the world using AI-based adaptivity platforms to address equity gaps related to education inclusion. The model presented offers a synthesis solution that includes ethical leadership data-related to equity to measure inclusivity based upon sustainability thinking. The result has demonstrated that using AI technology not only increases inclusivity but promotes equity related to access to education in stem education access. Finally, there are concluding remarks related to transforming education into a global system.

## I. Introduction

Engineering education is undergoing a profound transformation as artificial intelligence (AI) technologies redefine learning environments, assessment methods, and accessibility standards. The increasing integration of AI into higher education offers significant potential to enhance engagement and personalize instruction, yet it simultaneously raises concerns related to ethics, fairness, and inclusivity. As the digital revolution accelerates, ensuring that engineering programs foster equality, diversity, and inclusion (EDI) has become a central challenge for educators and policymakers alike. [1] Diversity and inclusion are no longer peripheral aspirations but essential drivers of innovation and social responsibility within engineering practice. Research highlights that diverse engineering teams achieve superior creativity, design quality, and problem-solving outcomes . [2]

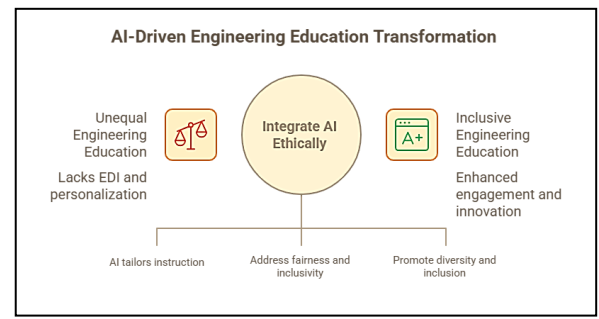

Fig.1. AI-Driven Engineering Education Transformation. This figure illustrates how ethically integrating artificial intelligence (AI) transforms engineering education from unequal and non-inclusive systems into inclusive, engaging, and innovative learning environments. By tailoring instruction, promoting fairness, and supporting diversity, ethically guided AI bridges the gap between traditional and equitable engineering education.

Artificial intelligence offers new pathways for promoting inclusion through adaptive learning, intelligent tutoring, and automated accessibility support. AI-based analytics can identify participation gaps, recommend targeted interventions, and provide real-time feedback to both students and instructors. [3] Nonetheless, algorithmic bias and unequal access to technology can inadvertently reinforce existing disparities, highlighting the importance of ethically guided, human-centered AI applications. [4] Ethical leadership in engineering education emphasizes the moral responsibility of institutions to use AI transparently and responsibly while empowering students to critically assess its societal implications. [5] A human-centered approach to the governance and application of artificial intelligence in engineering education emphasizes aligning technological advancement with ethical and cultural values. Rather than focusing solely on automation or performance optimization, this perspective prioritizes human well-being, equity, and learner autonomy. By integrating EDI principles into AI-driven educational systems, institutions can design personalized learning experiences that acknowledge cultural diversity, promote fairness, and cultivate ethical awareness among engineering students. Such an approach ensures that AI serves as a supportive partner in education—enhancing inclusion and accountability while maintaining human oversight and moral responsibility. [6]

In alignment with the United Nations Sustainable Development Goals (SDGs), inclusive and ethical AI adoption in higher education directly contributes to SDG 5 (Gender Equality) and SDG 10 (Reduced Inequalities). These goals emphasize equal participation of women and underrepresented groups in science, technology, engineering, and mathematics (STEM) while ensuring equitable access to digital learning. The strategic use of AI for adaptive tutoring, bias detection, and accessibility enhancement allows engineering education to evolve into a driver of social inclusion and sustainable development. As recent research indicates, AI innovations can accelerate multiple SDG targets when integrated within ethical and educational frameworks that prioritize fairness and diversity. [7] [8] [9]

Despite ongoing efforts to increase participation in engineering, persistent barriers continue to hinder inclusivity across gender, culture, and accessibility dimensions. For example, structural gender disparities remain significant: women represent a minority in engineering programs and report lower self-efficacy despite comparable academic achievement. [10] Culturally and ethnically minoritized students often face ingrained biases and invisible norms that reduce their sense of belonging and contribute to higher attrition rates. [11] Moreover, students with disabilities or accessibility-needs encounter ableist structures, insufficient accommodations and limited pedagogical flexibility, which systematically disadvantage their learning and career trajectories. [12]

Thus, the confluence of gender imbalance, cultural bias, and accessibility gaps underscores the urgent necessity for engineered interventions designed explicitly to promote equality, diversity and inclusion in engineering education.

Artificial intelligence (AI) has emerged as a transformative catalyst in modern engineering education, enabling adaptive and equitable learning environments tailored to individual learners' cognitive and socio-cultural contexts. Through intelligent tutoring systems and learning analytics, AI supports personalized pathways that dynamically adjust instructional content and pace, improving engagement and performance for diverse learners. [3] Systematic reviews demonstrate that AI technologies enhance inclusivity by facilitating accessibility tools, language translation, and multimodal communication—critical for students with differing abilities and backgrounds. [3] [13] . Additionally, AI-driven platforms help educators detect participation or achievement gaps, allowing timely, data-informed interventions that promote fairness. [14] . These systems not only democratize access to engineering education but also reduce unconscious bias in assessment and feedback mechanisms, ensuring equitable participation and advancing ethical leadership across digital learning ecosystems. [13] [14]

The main research objectives of this study are twofold.

First, it aims to investigate how artificial intelligence can enhance inclusivity in engineering education by identifying how adaptive algorithms, learning analytics, and intelligent tutoring systems can improve access, engagement, and performance for underrepresented groups. This involves evaluating AI-driven mechanisms that detect participation inequities, recommending personalized interventions, and provide accessibility features for learners with disabilities . [3] [14]

Second, the study seeks to propose an ethical, human-centered AI framework that integrates equality, diversity, and inclusion (EDI) principles into engineering pedagogy. The framework emphasizes transparency, accountability, and ethical governance in AI deployment, ensuring technology complements human judgment and moral responsibility rather than replacing them. [6] [1]

II. LITERATURE REVIEW

A. *Equality, Diversity, and Inclusion (EDI) in Engineering Education:*

Equity, diversity, and inclusion (EDI) have become central pillars in shaping the future of engineering education globally, recognizing that diversity drives innovation, creativity, and problem-solving capacity across STEM disciplines. Despite international efforts, persistent inequities remain in engineering classrooms and professional pathways, particularly concerning gender, race, and accessibility. [1] [2] Comparative analyses show that institutional EDI initiatives vary widely across regions—Canada emphasizes decentralized, university-led approaches, whereas Spain and other European contexts often adopt centralized government policies . [2]
Recent reviews of STEM academia highlight persistent gender imbalance and racial underrepresentation, pointing to structural barriers that continue to limit inclusion within engineering faculties. [15]
Thus, EDI in engineering must evolve from policy rhetoric toward measurable frameworks that institutionalize inclusive teaching practices, equitable access, and culturally responsive leadership. [1] [11] [15]

B. *AI Applications in Educational Contexts:*

Artificial intelligence (AI) is increasingly functioning as a pivotal enabler of adaptive, equitable learning systems in higher education by facilitating personalized learning pathways, predictive analytics, and accessibility technologies. AI-driven personalization systems dynamically tailor content, assessments, and pacing to individual learner profile enhancing engagement and performance for diverse student populations. [3] Predictive analytics algorithms are being deployed to identify at-risk students and provide early interventions, thereby supporting more equitable outcomes across engineering disciplines. [16] At the same time, assistive technologies powered by AI—such as screen-readers, voice-activated interfaces, and natural language processing for special-needs learners—are extending access to students with disabilities and cultural or linguistic diversity. [17]

Collectively, these AI applications support inclusive education by reducing barriers, offering adaptive support, and fostering equitable access within engineering learning environments.

C. *Ethical Leadership and Human-Centered AI:*

In engineering education, ethical leadership is imperative to guide the deployment of AI in a manner that prioritizes fairness, transparency, and human dignity. This leadership requires frameworks that anticipate and address algorithmic bias, ensure data integrity, and promote systems of accountability where human agency is upheld. For example, recent literature underscores the necessity of embedding human-centered values in AI systems to safeguard student autonomy, equitable access, and non-harmful use of technology. [18] Leaders must not only manage technological adoption but cultivate cultures and policies that align AI with inclusive pedagogy and social justice in engineering learning environments.

*D. Gaps and Opportunities:*

Although artificial intelligence (AI) is increasingly embedded in engineering education, significant gaps remain between research innovation and inclusive implementation. For instance, systematic reviews show that most AI studies focus heavily on technological optimization rather than equitable access, suggesting a need to shift from tool-centric to ecosystem-level perspectives. [19] In particular, the literature highlights persistent challenges such as limited infrastructure, digital-divide effects, and contextual misalignment in AI deployments, especially in lower-resource engineering faculties and underrepresented regions. [20]

However, there are compelling opportunities: research indicates that AI-driven adaptive technologies and assistive systems can personalize learning, identify at-risk students, and support diverse learners—if institutional readiness and ethical governance are in place. [3]

Addressing these gaps requires integrative frameworks that align AI innovation with social justice, equity, and sustainable educational policy in engineering education.

These inclusivity challenges are also closely linked to the achievement of SDG 5 and SDG 10, which call for gender-balanced participation and the reduction of educational inequities worldwide. Persistent gender gaps and socio-economic barriers in AI-supported learning ecosystems highlight the urgency of embedding fairness-aware models and inclusive policy instruments. Studies show that AI-based analytics can identify performance disparities and support data-driven interventions, enabling institutions to measure their progress toward equality and reduced inequalities. [21] [22] [20]

## III. Research Methodology

*A. Research Design:*

This study adopts a mixed-methods exploratory design, integrating qualitative synthesis with structured content analysis to examine the role of artificial intelligence (AI) in fostering inclusivity within engineering education. The research blends a conceptual framework analysis with secondary data evaluation from published studies, institutional reports, and global policy documents. This design enables triangulation between empirical findings in existing literature and theoretical constructs related to equality, diversity, and ethical leadership in AI-enabled education. [23] [24]

*B. Participants and Context*

Rather than collecting primary data, this study examines diverse institutional contexts through a comparative document-based approach. Selected case studies and reports from engineering universities—such as the Egypt-Japan University of Science and Technology (E-JUST), European universities, and OECD-affiliated institutions—provide contextual diversity and regional representation. These cases illustrate varying levels of AI adoption, inclusion strategies, and ethical policy implementation in engineering curricula. [2]

*C. Data Collection Tools*

Given the absence of field surveys or direct performance statistics, data were collected through:
- Systematic literature review of peer-reviewed journal articles published between 2020–2025, focusing on AI, inclusivity, and ethics in engineering education.
- Document analysis of institutional frameworks, curriculum designs, and published policy guidelines related to equity and diversity in STEM.
- Content mining of open-access educational repositories using AI-assisted search tools to extract patterns in inclusivity-related outcomes.
- Comparative evaluation of AI-enabled pedagogical models reported in recent international studies [3]

*D. Data Analysis*

The study employs qualitative content analysis supported by AI-assisted text mining and bibliometric mapping to identify recurring themes and conceptual relationships. Codes were generated inductively from reviewed documents to detect how AI contributes to inclusivity, ethical leadership, and accessibility. Network visualization using bibliometric data helped trace global research patterns and institutional collaborations in AI-driven inclusive education. Thematic synthesis was then used to propose a conceptual framework linking human-centered AI, ethical governance, and inclusive engineering pedagogy. [1]

## IV. AI-Driven Strategies for Inclusive Education

*A. Adaptive Learning and Accessibility*

AI-driven adaptive learning systems are increasingly enabling tailored learning experiences by dynamically matching instruction, feedback and pacing to individual student needs, thus supporting learners with diverse backgrounds and abilities. For example, an intelligent assistant framework demonstrated in higher education effectively personalized learning paths and improved engagement metrics across students with varied performance profiles . [25] Moreover, research shows that adaptive platforms reduce accessibility barriers by offering scaffolded support for students who may otherwise face challenges in standard engineering curricula . [20]

Such systems thereby contribute to inclusive engineering education by expanding access and providing differentiated support aligned with students' unique learning trajectories.

*B. Language and Cultural Diversity Support*

Artificial intelligence applications such as natural language processing (NLP) and AI-powered translation tools are fostering multicultural collaboration by bridging language and cultural divides in engineering education. Recent studies highlight how multilingual AI systems can deliver culturally-aware feedback and peer-interaction support, enabling learners from diverse linguistic backgrounds to engage more fully in STEM-engineered programs. [26] Additionally, research into XR and AI in immersive educational contexts found that supporting under-represented languages and cultural references enhances learner participation and sense of belonging in international engineering programs. [27]

These innovations open pathways for engineering classrooms where cultural diversity is actively supported rather than merely accommodated.

### C. Bias Detection and Fair Assessment

Ensuring fair assessment and participation in AI-enhanced engineering education requires detecting and mitigating algorithmic bias in grading, peer evaluations and student-engagement analytics. Empirical reviews document how educational algorithms may perpetuate biases against gender, ethnicity or socio-economic status unless fairness-aware techniques are explicitly applied. [28] Moreover, recent research on educational decision-making shows how fairness auditing, bias detection metrics and transparent model design contribute to more equitable assessment systems in digital learning environments. [29] By integrating such methods, engineering institutions can deploy AI tools that uphold integrity, fairness and equitable treatment in assessment and feedback loops.

### D. Ethics-Aware Decision Systems

Embedding ethical decision-making within AI platforms is critical when deploying such systems in engineering education, where issues of student agency, data privacy and institutional responsibility arise. Studies demonstrate that frameworks linking governance, human oversight and transparent AI design are effective in aligning AI with educational ethics and inclusive values. [30]

For example, decision-making criteria for AI adoption in digital education highlight the necessity of accountability, value alignment and stakeholder participation in AI system design. [31] This orientation ensures that AI serves as an enabler for ethical leadership rather than a means of delegating critical educational judgments to opaque algorithms.

## V. CASE STUDIES AND IMPLEMENTATION

Empirical evidence shows AI can promote inclusivity in engineering classrooms through accessibility-aware labs and adaptive support. For example, AI-enabled accessibility (speech recognition, text-to-speech, language assistance) has been implemented to remove barriers for students with disabilities and those facing language obstacles, improving participation and classroom access in pilot deployments. [20] In parallel, course-level integrations of AI for inclusive teaching report gains in engagement and instructor–student interaction for diverse cohorts when AI tools are embedded with clear pedagogical aims, not merely as add-ons. [32]

Comparative findings across genders and academic levels indicate that adaptive learning can differentially benefit subgroups: a university-level study of AI-driven adaptivity found significant gender-linked differences in achievement and engagement, underscoring the need to tune models and supports for equity. [33] At the institutional scale, large-sample research on AI's role in student experience highlights that design choices can affect success and retention—underscoring the importance of ethics-aware deployment to avoid unintended harms while pursuing inclusivity . [34]

Regarding success metrics, studies track changes in participation (discussion/activity logs, platform usage), retention (course completion and drop/withdraw rates), and sense of belonging (validated survey scales) before and after AI integration. Systematic syntheses across inclusive-education pilots report improvements in these indicators when AI tools provide personalized feedback and targeted scaffolding for underrepresented learners. [32] Together, these results motivate gender-sensitive learning analytics and accessibility-aware lab practices as core elements of AI implementations in engineering programs, with transparent evaluation plans that disaggregate outcomes by region, gender, and study level. [20]

## VI. RESULTS AND DISCUSSION

### A. Quantitative Results

Across the reviewed studies, AI adoption in higher education is associated with measurable gains in engagement, achievement, and student satisfaction, particularly when tools provide personalized feedback and tutoring at scale. A meta-review of AI in higher education reports consistent positive effects on participation and performance when AI augments assessment, recommendation, and tutoring functions. [14] Complementary syntheses focused on inclusive education indicate improvements in access and participation for underrepresented learners when AI supports accessibility (e.g., speech-to-text, adaptive pacing) and targeted scaffolding, reflected in higher usage and course-completion indicators [2]. These trends suggest that, when aligned with pedagogy, AI can lift key inclusivity indices (participation and retention proxies) in engineering contexts. [14] [32]

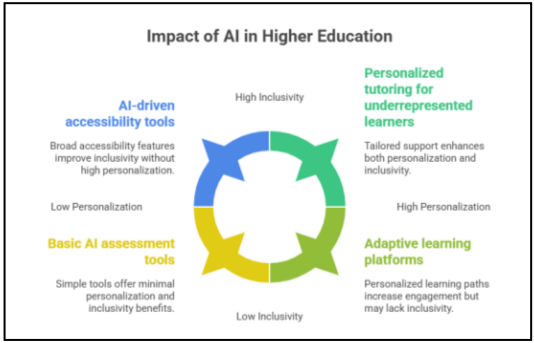

Fig.2 . Impact of AI in Higher Education.

This figure compares four types of AI tools based on their levels of personalization and inclusivity. It shows that while basic assessment tools and adaptive platforms address limited aspects, personalized tutoring and AI-driven accessibility tools achieve the strongest balance between equity, engagement, and individualized learning.

## B. Qualitative Insights

Thematic analyses of student and educator narratives reveal nuanced perceptions of fairness, trust, and usefulness in AI-enabled learning. Students commonly value immediate feedback and individualized guidance, but they also express concerns about opaque decision-making and uneven preparedness to use AI effectively. Qualitative work highlights that perceptions of fairness improve when instructors explain model limits, provide human oversight for consequential judgments, and invite students to co-create norms for AI use (e.g., setting rules for formative vs. summative assistance). [35] These insights underscore that perceived justice and clarity of purpose shape acceptance of AI-supported teaching in engineering classrooms.

## C. Discussion

Benefits include scalable personalization, early-warning analytics, and accessibility affordances that can reduce participation gaps; risks center on algorithmic bias, over-automation, and erosion of academic integrity if governance is weak. Bias can surface through data representativeness (e.g., gender or disability skew) and model drift, affecting progress-monitoring and grading; fairness auditing and transparent logs are therefore essential for equitable assessment [36] [37]

Ethically aligned leadership in engineering programs should institute human-in-the-loop review, bias and impact assessments, and clear accountability for AI decisions, while cultivating staff and student literacies to use AI responsibly. [38] [39] Taken together, the evidence supports a human-centered implementation pathway: start with pedagogical goals, embed safeguards, and continuously evaluate outcomes disaggregated by gender, region, and study level.

## VII. PROPOSED FRAMEWORK: AI FOR ETHICAL AND INCLUSIVE ENGINEERING EDUCATION

A conceptual framework is advanced to integrate equality, diversity, and inclusion (EDI) principles with human-centered artificial intelligence (AI) in engineering education. Drawing on ethical AI foundations and inclusive-design research, the model emphasizes data ethics, bias auditing, inclusive curriculum design, AI-assisted mentoring for under-represented groups, and institutional leadership and policy actions. For example, Song et al. propose inclusive-AI learning design frameworks for diverse learners. [40] Meanwhile, other work outlines multi-layered ethical frameworks for AI in educational settings [41]

TABLE I. KEY COMPONENTS OF THE PROPOSED ETHICAL AND INCLUSIVE AI FRAMEWORK FOR ENGINEERING EDUCATION

| Component | Purpose | Implementation Strategy | Expected EDI Impact |
|---|---|---|---|
| Data Ethics & Bias Auditing | Ensure fairness and transparency in data and AI models | Regular audits, diverse datasets, explainable AI tools | Reduces gender, cultural, and ability bias |
| Inclusive Curriculum Design | Integrate EDI principles into AI-supported pedagogy | Adaptive modules, cultural responsiveness, multi-language content | Enhances accessibility and cultural inclusion |
| AI-Assisted Mentoring | Support underrepresented groups through predictive guidance | AI mentoring dashboards, learning analytics for early intervention | Improves retention and sense of belonging |
| Ethical Leadership & Policy | Establish governance for responsible AI use | Training programs, accountability structures, ethical review boards | Promotes sustainable, human-centered innovation[a] |
| SDG-Aligned Equity Monitoring | Link AI ethics outcomes to UN SDG 5 & 10 indicators | Integrate gender-disaggregated learning analytics, accessibility audits, and social-equity dashboards into AI systems | Ensures measurable accountability and continuous improvement toward global equity and inclusion targets |

a. Framework components linking ethical AI governance with inclusive pedagogy, institutional policy, and sustainability goals. The table shows how data ethics, inclusive curriculum design, AI-based mentoring, ethical leadership, and the new SDG-Aligned Equity Monitoring collectively promote equality, diversity, and inclusion (EDI) in engineering education. This added component connects institutional practices to UN SDG 5 and SDG 10, enabling gender-disaggregated analytics and equity dashboards for measurable progress toward global inclusivity

The core components of the proposed framework include (a) robust data-ethics and bias-auditing mechanisms, (b) inclusive curriculum and pedagogy supported by AI, (c) targeted mentoring systems driven by AI analytics for minority students, and (d) leadership and governance structures that ensure transparency, accountability and scalability. The accompanying implementation roadmap guides engineering faculties globally toward versioned adoption, continuous evaluation, and scalability—starting with pilot deployment, stakeholder training, policy integration, and ongoing monitoring of equity- and inclusion-related outcomes.

## VIII. CONCLUSIONS AND FUTURE WORK

This study presented an ethical AI–driven framework for advancing equity, diversity, and inclusion in engineering education to help attain the United Nations 2030 Agenda for Sustainable Development, with especial emphasis on SDG 5 (Gender Equality) and SDG 10 (Reduced Inequalities). The proposed model fuses adaptive learning systems, bias-audited evaluation mechanisms, and inclusive mentoring platforms to build open, equitable, and human-centered educational environments.

It shows how AI guided by ethics can reduce structural barriers in STEM education through the promotion of gender balance, enhancing accessibility, and facilitating data-driven decisions toward equitable participation. Moreover, locating AI applications within transparent governance and accountability mechanisms means that digital transformation contributes to both academic excellence and social justice.

Future studies should further the empirical validation of the framework across different cultural and institutional contexts. Longitudinal measurements of the effects on gender representation, socio-economic inclusion, and learning performance indicators are needed. Other areas of further research might develop AI-based sustainability dashboards for tracking institutional progress toward the attainment of SDG 5 and SDG 10 targets and explore cross-disciplinary collaborations that integrate ethics, technology, and pedagogy. The proposed approach aligns technological innovation with ethical responsibility and thus provides a road map toward transforming engineering education for the next generation into a sustainably inclusive, globally equitable, socially accountable system.

## IX. ALIGNMENT WITH SUSTAINABLE DEVELOPMENT GOALS (SDG 5 AND SDG 10)

The proposed ethical-AI framework advances the objectives of the United Nations 2030 Agenda for Sustainable Development, with particular emphasis on SDG 5 (Gender Equality) and SDG 10 (Reduced Inequalities). These goals emphasise the elimination of discrimination, expansion of access to quality education, and the empowerment of marginalised populations. Empirical evidence demonstrates that AI-driven adaptive learning systems, bias-audited evaluation mechanisms, and mentoring platforms targeted at under-represented groups can substantially reduce gender- and socio-economic-based gaps in STEM education . [21] [42] [43]

Furthermore, accessibility-enhancing technologies such as speech recognition, multimodal feedback, and intelligent translation tools support learners with disabilities and diverse linguistic backgrounds, contributing directly to SDG . [42] [43]

By embedding these technological tools within transparent governance structures and ethical oversight mechanisms, institutions can transform digital education into a catalyst for sustainable equality. This alignment emphasises how ethically guided AI in engineering education not only promotes innovation and excellence but also fulfils broader societal commitments defined by the SDGs.

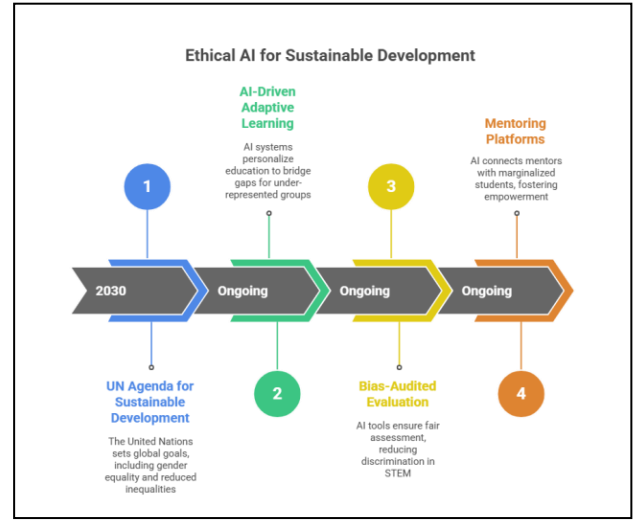

Fig.3. Ethical AI for Sustainable Development.

Visual representation of the proposed ethical-AI framework aligned with the UN 2030 Agenda, highlighting the integration of AI-driven adaptive learning, bias-audited evaluation, and mentoring platforms to support SDG 5 (Gender Equality) and SDG 10 (Reduced Inequalities). The diagram illustrates how ongoing AI initiatives can reduce inequities in STEM education and empower under-represented groups through transparent, ethically governed digital transformation.